\documentclass{iaus}
\usepackage{graphicx}

\title[SWAS observations of comet 9P/Tempel 1 and Deep Impact] 
{SWAS observations of comet 9P/Tempel 1 and Deep Impact}

\author[Bensch et al.]   
{Frank Bensch$^{1,2}$ 
 Gary J. Melnick$^2$ David A. Neufeld$^3$ \break 
 Martin Harwit$^4$ Ronald L. Snell$^5$ \and Brian M. Patten$^2$}

\affiliation{$^1$Radioastronomisches Institut, Universit\"at Bonn,
Auf dem H\"ugel 71, 53121 Bonn, Germany
\\[\affilskip]
$^2$Harvard-Smithsonian Center for Astrophysics, 60 Garden Street, Cambridge,
MA 02138, USA \\ [\affilskip]
$^3$Department of Physics and Astronomy, Johns Hopkins University, 
Baltimore, MD, USA \\ [\affilskip]
$^4$511 H street, SW, Washington, DC, USA; 
also Cornell University \\ [\affilskip]
$^5$Department of Stronomy, University of Massachusetts, Amherst, 
MA 01003, USA 
}

\pubyear{2005}
\volume{231}  
\pagerange{1--2}
\date{15 September 2005 and in revised form ??}
\jname{Astrochemistry - Recent Successes and Current Challenges}

\editors{D.C. Lis, G.A. Blake \& E. Herbst, eds.}
\begin{document}

\maketitle

\begin{abstract}
On 4 July 2005 at 1:52 UT the Deep Impact mission successfully completed
its goal to hit the nucleus of 9P/Tempel\,1 with an impactor, forming a crater 
on the nucleus and ejecting material into the coma of the comet 
(\cite{AHearn2005}). 
The 370 kg impactor collided with the sunlit side of the nucleus with a 
relative velocity of 10.2 km/s. NASA's Submillimeter Wave Astronomy Satellite 
(SWAS) observed the $1_{10}-1_{01}$ ortho-water ground-state rotational 
transition in comet 9P/Tempel 1 before, during, and after the impact.  
No excess emission from the impact was detected by SWAS. However,
the water production rate of the comet showed large natural variations 
of more than a factor of three during the weeks before the impact.
\end{abstract}

\begin{figure}
\begin{centering}
 \includegraphics[width=40mm,angle=-90]{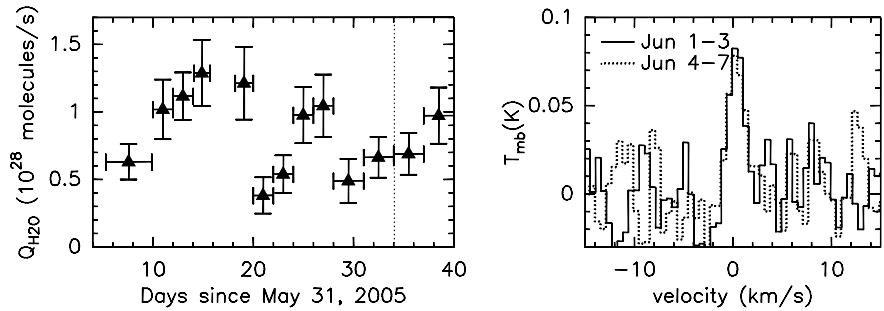}
  \caption{Left panel: Water production rate of comet 
9P/Tempel 1, derived from SWAS observations of the  $1_{10}-1_{01}$ 
transition of o-H$_2 ^{16}$O. The vertical line gives the impact time.
Right panel: SWAS spectrum before and after the 
impact. Each spectrum is a co-average of 3 days.}
\label{q-vs-time}
\end{centering}
\end{figure}

SWAS is a complete space-borne radio observatory (\cite{Melnick2000}), 
capable of observing the 556.9\,GHz transition of 
ortho H$_2 ^{16}$O, among other species, with a velocity resolution of 
1\,km\,s$^{-1}$ and a beam size of $3.3\times4.5$\,arcminutes. SWAS began 
near-daily monitoring observations of comet 9P/Tempel 1 on 5 June 2005, 
and the observations were continued post-impact  until 1 September 2005.
The total water production rate $Q$ of the comet is determined from the 
velocity-integrated intensity of the emission detected in the
SWAS spectra (typically being 2 to 3-day co-adds) and employing the 
radiative transfer model for water line emission in comets 
by \cite{Bensch2004}. 
The left panel of Fig.\ \ref{q-vs-time}
shows the SWAS-measured water production rate for the observations 
made from 5 June through 9 July. The cometary 
activity traced by the water evaporation rate varies by more than a factor 
of three, from $3.8\times10^{27}$ to $12.9\times10^{27}$\,
molecules per second (115 to 385 kg\,s$^{-1}$, respectively). 

No statistically significant increase in the water line emission was detected 
by SWAS immediately following the impact. The average velocity-integrated 
intensity during the three days after the impact is 0.16$\pm$0.04\,K\,km/s,
virtually identical to the average intensity measured for the three
days before impact (Fig.\ \ref{q-vs-time}, right panel). 
This corresponds to a total water production rate 
of $(6.6\pm1.5)\times 10^{27}$\,s$^{-1}$. 

\begin{figure}
\begin{centering}
 \includegraphics[width=40mm,angle=-90]{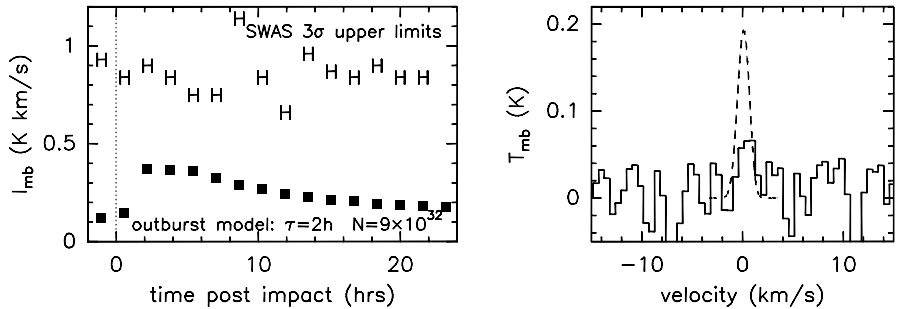}
  \caption{Left: The velocity-integrated intensity calculated by
an outburst model with $\tau=2$\,hrs and $N=9\times 10^{32}$ 
is compared to the 3$\sigma$ upper limit derived from the SWAS spectra
observed during each orbital segment of 35\,min duration (error bars). 
No emission is detected in individual segments, consistent with the low 
$Q_{\rm H_2O}$ measured before impact. Right panel: 
the weighted average of line profile for this outburst model 
(dashed line) is compared to the SWAS-measured spectrum. The individual 
spectra are weighted by the square of integrated intensity predicted by the 
model for each orbital segment and the inverse square of the
rms noise in the spectrum.}
\label{outburst-model}
\end{centering}
\end{figure}

In order to derive an upper limit on the water released by the impact,
we extended the radiation transfer model to include comets where the 
water production rate is time-dependent.
The total water production rate is replaced by $Q=Q_q + Q_b(t)$, where
$Q_q$ is the constant (quiescent) component. $Q_b(t)$ is the
time-variable component, assumed to be a box-car function
for the present simulations. In this case the water production rate is 
elevated (but constant) for the duration $\tau$ of the outburst and 
returns to the pre-outburst level afterwards. The total number of water 
molecules released by the outburst is $N=Q_{\rm b}\times \tau$. An example 
for the time evolution of the velocity-integrated intensity derived with this
outburst model is shown in 
Fig.\ \ref{outburst-model}.
Preliminary model results for the SWAS observations up to 24\,hrs after
the impact 
give a 3$\sigma$ upper limit of $\sim$9$\times 10^{32}$ molecules 
($2.7\times 10^4$ tons) for the vaporized water.
This result
is not very sensitive to the assumed duration of the outburst, as 
a similar upper limit is obtained for models with 
$\tau=$0.5, 2, 8 and 16\,hrs. In addition, a doubling of the
water production rate can be excluded at a $3\sigma$ confidence level
if we assume that the impact created a new, permanently active area on the 
comet nucleus. Results from
Odin observations of the ortho-water $1_{10}-1_{01}$ transition made during the
Deep Impact event are presented by D.\ Despois (this volume).

\begin{acknowledgments}
Support of the SWAS mission is provided by NASA through SWAS contract 
NAS5-30702. Additional support for the SWAS 9P/Tempel 1
observing campaign was obtained by the Deep Impact team. The help and 
cooperation of NASA HQ is acknowledged.
\end{acknowledgments}

%
\end{document}